\documentclass[conference]{IEEEtran}
\IEEEoverridecommandlockouts
\usepackage[T1]{fontenc}
\usepackage[utf8]{inputenc}
\usepackage{cite}
\usepackage{amsmath,amssymb}
\usepackage{graphicx}
\usepackage{booktabs}
\usepackage{array}
\usepackage{multirow}

\usepackage{microtype}
\usepackage[hidelinks]{hyperref}
\usepackage{balance}
\usepackage[dvipsnames,table,xcdraw]{xcolor}
\usepackage{xurl}

\graphicspath{{figures/}}
\hypersetup{
 pdftitle={A Hybrid Sionna RT-Based NTN Channel Simulation Framework: Software Architecture, Statistical Augmentation, and LSP Validation},
 pdfauthor={Yu-Chien Lin, Cheng-Yi Chang, Yen-Cheng Lai, Ming-Chun Lee, and Ta-Sung Lee},
 pdfsubject={IEEE conference manuscript},
 pdfkeywords={NTN, Sionna RT, hybrid channel model, ray tracing, large-scale parameters, OpenNTN}
}
\newcommand{\RT}{\mathrm{RT}}
\newcommand{\STAT}{\mathrm{stat}}
\newcommand{\dd}{\mathrm{d}}
\newcommand{\code}[1]{\texttt{#1}}

\newcommand{\figorplaceholder}[3]{%
  \IfFileExists{figures/#2}{\includegraphics[width=#1]{#2}}{%
    \fbox{\parbox[c][#3][c]{0.90\columnwidth}{\centering Missing figure: \\ \texttt{\detokenize{#2}}}}%
  }%
}

\begin{document}

\title{An Open-Source Hybrid Ray-Tracing and 3GPP Channel Simulator for Non-Terrestrial Networks}

\author{
Yu-Chien Lin,
Cheng-Yi Chang,
Yen-Cheng Lai,
Ming-Chun Lee,
and Ta-Sung Lee
\thanks{
Institute of Communications Engineering,
National Yang Ming Chiao Tung University, Hsinchu 300093, Taiwan. Email:
heabajoshua@gmail.com, bensonxlol@gmail.com
alvin92828tw.ee14@nycu.edu.tw, mingchunlee@nycu.edu.tw, tslee@nycu.edu.tw.
}
}

\maketitle

\begin{abstract}
This paper proposes a hybrid channel simulation framework that combines site-specific ray tracing (RT) with 3GPP-compliant statistical channel simulation for non-terrestrial networks (NTNs). The proposed approach uses RT to generate dominant propagation paths and augments them with statistically generated paths and subpaths to provide richer multipath realizations under practical simulation complexity. To provide a complete simulation framework, we integrate Sionna RT, 3GPP, and hybrid RT-3GPP simulators within a unified workflow. Simulation results based on key channel statistics and power delay profiles show that the proposed method preserves site-specific propagation characteristics while providing sufficiently rich multipath components, thereby addressing the complementary limitations of conventional RT-based and purely statistical NTN channel simulators. The developed simulator is also released as an open-source project to support reproducibility and further NTN research.
\end{abstract}


\section{Introduction}\label{sec:intro}

To support ubiquitous connectivity in remote and infrastructure-limited environments, the use of non-terrestrial networks (NTNs) to support existing terrestrial networks has been considered \cite{6Gmobile_1,huang2024joint}. In particular, the integration of ground base stations (BSs) and low earth orbit (LEO) satellites has emerged as a paradigm for next-generation wireless systems \cite{6Gmobile_2}. Nevertheless, the propagation conditions of NTNs differ
substantially from those encountered in conventional terrestrial systems. Especially, satellite links exhibit large propagation distances, elevation-dependent
blockage and fading, and rapidly varying delay and Doppler caused by satellite motions. This thus calls for different simulation methodologies and modeling for NTN channels, where two major types of channel modeling and simulation methodologies, i.e., statistical channel simulation and ray-tracing (RT) simulation, are considered \cite{kurner2026review,wu2026implementation}. 

One of the most representative statistical channel simulation methods is reported in 3GPP TR 38.811~\cite{tr38811}, where considerable effort has been devoted to this standardized NTN channel simulation~\cite{tr38811}. Then, \cite{sandri2023} implemented the TR 38.811 NTN channel model in ns-3 with additional features and validated the implementation against the 3GPP calibration results. Similarly, OpenNTN~\cite{openntn} provides an open-source implementation of 3GPP-based NTN channel models within the NVIDIA Sionna ecosystem. In addition, MATLAB-based NTN simulation environments have been developed in~\cite{milovanovic2027} to support configurable link- and system-level evaluations based on standardized channel models. While these approaches provide efficient and reproducible
channel realizations over broad classes of NTN scenarios, their simulation results, however, are primarily scenario- and parameter-driven under statistical modeling rather than derived from the explicit geometry of a site-specific environment.

To simulate channels with site-specific results, RT-based methods have been discussed \cite{kurner2026review}. In particular, Sionna RT provides an RT-based simulation platform within the NVIDIA ecosystem and supports GPU acceleration for efficient channel simulation~\cite{sionna2023,sionna2025}. In NTN scenarios, however, very large satellite-ground distances together with finite RT sampling can result in some receiver positions having only a small number of detected multipath components. Such sparse realizations might lead to skewed delay spread (DS), excessively large K-factor, or poorly conditioned angular statistics. To enable a hybrid version of the statistical and RT-based simulation methods, 3GPP TR 38.901~\cite{tr38901} establishes the general concept and high-level methodology of the map-based hybrid modeling, in which deterministic RT on a digitized environment is supplemented by stochastic channel components. Hence, this hybrid model is suitable for complementing the existing drawback of the RT-based simulation for NTN scenarios. However, the implementation details have not been discussed in~\cite{tr38901}.

To accelerate the simulator and enable hardware-in-the-loop architecture for real-time testing, emulation methods have been developed and discussed~\cite{sun2026}. For example, \cite{muro2024} developed an OpenAirInterface- and Kubernetes-based 5G-NTN emulation platform, in which NTN characteristics are emulated at the network level while abstracting the detailed radio-link implementation. In addition, \cite{fang2026} proposed a hardware-in-the-loop platform that maps scenario information to real-time emulation of propagation delay, Doppler shift, fading, and path loss for 5G-Advanced NTN experiments. While these approaches enable real-time channel generation and emulation, hybrid methodologies that combine RT-based and statistical channel modeling remain unexplored.


To address the limitations of existing NTN channel simulators, where statistical models cannot provide site-specific channel realizations while RT-based methods may fail to generate sufficiently rich multipath components under feasible computational complexity, this paper proposes a simulation approach that integrates site-specific and statistical channel-modeling simulation methods within a unified and executable NTN channel-generation workflow. Specifically, we first develop a complete NTN channel simulation framework consisting of a site-specific simulator implemented using Sionna RT, a 3GPP statistical simulator, and the proposed hybrid RT-3GPP simulator. We then present the simulation approach and implementation details of the proposed hybrid RT-3GPP simulator, following the hybrid channel modeling architecture discussed in \cite{tr38901}. The key idea is to use the RT simulator to generate the dominant site-specific propagation paths, which serve as the main clusters. Based on these RT-generated paths, additional weaker paths and subpaths are statistically generated according to the statistical models adopted in 3GPP. In this manner, the proposed approach preserves the site-specific propagation characteristics captured by RT while enriching the channel with additional multipath components that may otherwise be missing because of the limited number of rays received by a receiver under practical RT simulation complexity. We validate the proposed simulator by comparing key channel statistics and power delay profiles (PDPs) obtained from the Sionna RT, 3GPP, and proposed hybrid RT-3GPP simulators. The results demonstrate that the proposed hybrid simulator can simultaneously preserve site-specific propagation characteristics and generate sufficiently rich and statistically meaningful multipath components. It therefore effectively addresses the complementary limitations of conventional RT-based and purely statistical NTN channel simulators. Finally, to facilitate reproducibility and further NTN research, the developed simulator has been released as an open-source project on GitHub.\footnote{GitHub repository: \url{https://github.com/Ollie-92/HybridChannel}.}

\section{Simulator Architecture and Simulation Flow}\label{sec:architecture}
\begin{figure*}[t]
\centering
\includegraphics[width=0.9\textwidth]{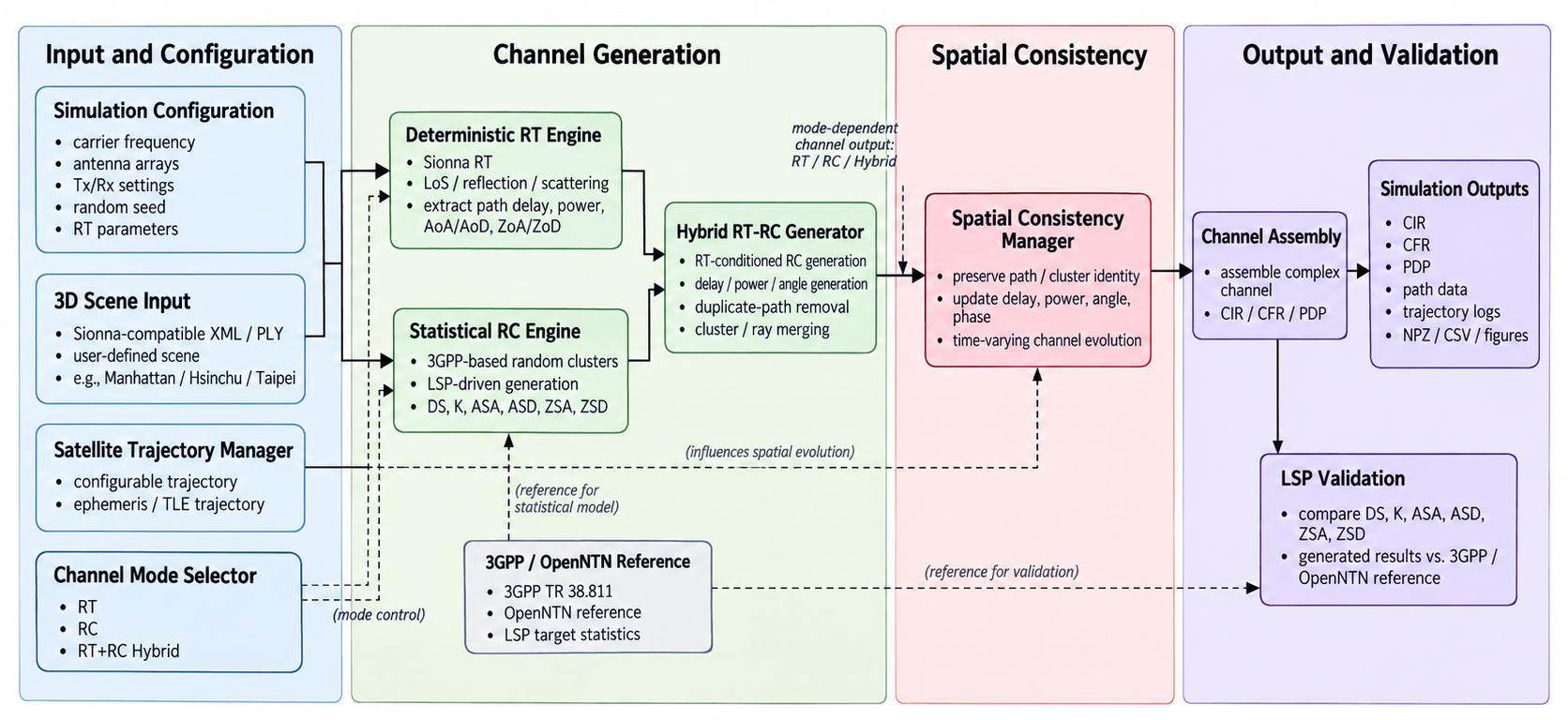}
\vspace{-10pt}
\caption{Block diagram of the proposed NTN channel simulation framework.}
\vspace{-5pt}
\label{fig:architecture}
\end{figure*}

This paper provides a simulation framework that unifies the 3GPP simulator, the Sionna RT-based simulator, and the proposed hybrid RT-3GPP simulator for channels in NTN networks. Fig.~\ref{fig:architecture} illustrates the overall simulation flow of the proposed framework. Starting from the input and configuration settings, the simulator first retrieves the context information for the simulation, which could include a Sionna-compatible scene, satellite positions from the TLE/Skyfield ephemeris, TX-RX configurations, simulation modes, etc. Then, the Sionna RT is executed if the RT-only or hybrid RT-3GPP simulator is adopted, and otherwise, only the 3GPP simulator is invoked. Then, after channels are generated, the spatial consistency is created to make sure channel correlations among time and space are appropriately established. After this, the channels are output for use.

\subsection{Scene and Ephemeris-Based Satellite Geometry}

While the simulator provides several default scenes, a user can customize its own scene by supplying a Sionna-compatible XML scene together with the referenced PLY meshes. Since the scene preprocessing could be conducted externally, the customized scene may not increase the runtime of the simulator. In this paper, the Manhattan scene is used as the default scene for discussion below.

Satellite motion is obtained from two-line element (TLE) data using Skyfield~\cite{skyfield2019}. The Skyfield \code{EarthSatellite} interface and a WGS84 observer are used to compute the time-varying satellite geometry, including azimuth, elevation, slant range, and range-rate-derived Doppler. The resulting ephemeris trajectory is mapped into the simulation geometry and applied at each time step. The same satellite position can therefore drive the deterministic RT calculation, the statistical channel evolution, and their combination.

\subsection{Channel Generation Modes}

\begin{figure*}[t]
\centering
\includegraphics[width=0.9\textwidth]
{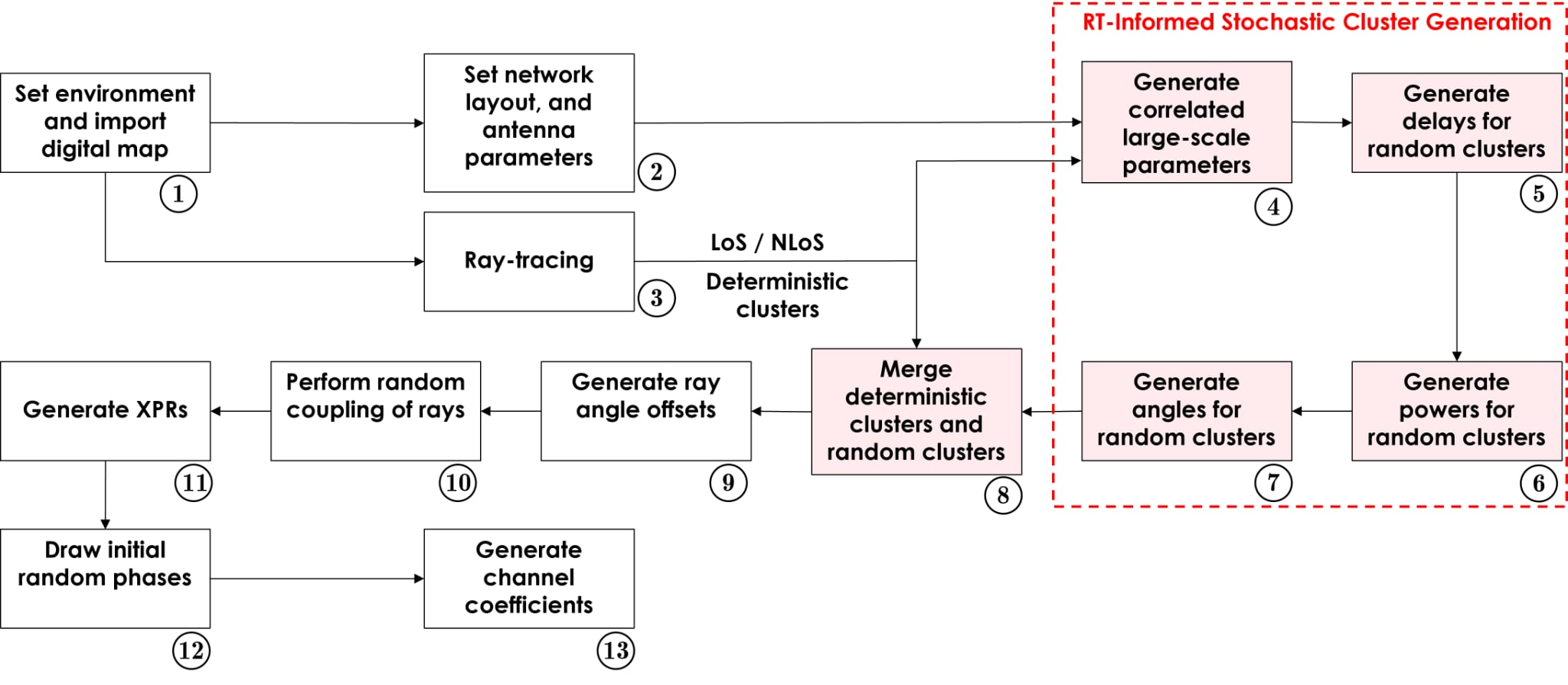}
\vspace{-10pt}
\caption{Illustration of hybrid RT-3GPP channel simulation
procedure.}
\vspace{-10pt}
\label{fig:hybrid_pipeline}
\end{figure*}

The framework provides configurable mode selection among Sionna RT, 3GPP statistical, and hybrid RT-3GPP channel generation methods. In
Sionna RT mode, the channel is constructed deterministically by using Sionna RT paths. In the 3GPP mode, the channel is generated using OpenNTN~\cite{openntn}, an open-source implementation of NTN channel models based on 3GPP TR 38.811~\cite{tr38811}. In the hybrid mode, the deterministic RT paths serve as site-specific anchors are first generated, and are then augmented by statistical multipath components generated from the 3GPP simulator. As illustrated in Fig.~\ref{fig:hybrid_pipeline}, the idea of the hybrid mode is to let the paths of RT serve as the site-specific main paths for the NTN channels. However, because of the large propagation distance between the satellite and ground users, weak multipath components are difficult to capture unless the simulation complexity is increased substantially. Hence, by treating the RT paths as the dominant clusters, weaker multipath components are then augmented using the subpaths generated by a 3GPP simulator. This thus helps capture the small-scale randomness that is difficult to generate using RT alone. Furthermore, some statistical clusters will also be augmented to enable richer multipath components. We note that while this methodology has been illustrated in 3GPP TR 38.901 \cite{tr38901}, the implementation details are not provided by it.

For time-varying simulations, the Sionna RT simulator updates deterministic paths at each satellite position. In the 3GP simulator,  spatial consistency for statistical channel realizations is ensured by approaches in~\cite{tr38901}. For the hybrid mode, the proposed frozen-scatterer mechanism maintains statistical cluster identities across satellite positions, while deterministic RT paths are updated from the ephemeris. The two components are then combined to form a time-varying hybrid channel, as will be detailed later in Sec.~\ref{sec:consistency}.





\section{Proposed Hybrid RT-3GPP Channel Simulator}\label{sec:hybrid}
This section details the proposed hybrid RT-3GPP channel generation procedure, where the downlink line-of-signt (LoS) case is discussed in this paper. However, the extension to non-LoS (NLoS) or uplink cases can follow a similar procedure. Fig.~\ref{fig:hybrid_pipeline} illustrates the complete
procedure of the proposed hybrid RT-3GPP channel simulator, which
includes deterministic RT path extraction, correlated large-scale parameter (LSP) sampling, statistical cluster generation, cluster
merging, subpath expansion, and channel coefficient construction.

\subsection{Site-Specific Multipath Components Generation}

To realize the hybrid simulator, we first need to determine the dominant paths of the channel. To this end, at each satellite position obtained from the ephemeris trajectory, the RT simulator is conducted once to determine whether the channel is in an LoS or NLoS scenario. If a physical LoS path is present, we classify the channel as an LoS scenario; otherwise, it is an NLoS scenario. This classification then determines whether the subsequent subpath generation follows the LoS or NLoS procedure of the 3GPP-compliant simulator. It should be noted that if the channel is categorized as with the LoS scenario, the path with smallest delay will be considered as a LoS path. Furthermore, for those paths whose delay is within 1 ns of the very first LoS path, they are also considered as LoS paths. Then, the remaining paths are considered as NLoS paths in the LoS scenario. In addition, for those RT paths whose power is below $25$ dB of the strongest path, they could be removed from the subsequent procedure to reduce complexity, as suggested in \cite{tr38901}. Each retained RT path $i$ can then be represented by using its parameter set, given as:
\begin{equation}
 \mathcal{P}^{\RT}_i=
 \left\{\tau_i^{\RT},P_i^{\RT},
 \phi_{i,\mathrm{AOA}},\phi_{i,\mathrm{AOD}},
 \theta_{i,\mathrm{ZOA}},\theta_{i,\mathrm{ZOD}}\right\},
 \label{eq:rt_path}
\end{equation}
where $\tau_i^{\RT}$, $P_i^{\RT}$, $\phi_{i,\mathrm{AOA}}$, $\phi_{i,\mathrm{AOD}}$, $\theta_{i,\mathrm{ZOA}}$, and $\theta_{i,\mathrm{ZOD}}$ are the excess delay, power, azimuth angle of departure (AOD), azimuth angle of arrival (AOA), zenith AOD, and zenith AOA for path $i$, respectively. Given the paths generated by the RT simulator, we then first augment some statistically-generated NLoS paths, and then augment each NLoS path with its associated subpaths to construct the complete set of multipath components of the channel. Note that for those LoS paths, they do not need to be augmented with subpaths. 


\subsection{Correlated LSP Sampling}
To generate augmented paths and sub-paths, the hybrid simulator first samples $6$ LSPs, i.e., $K$, DS, ASD, ASA, ZSD, and ZSA. Since these parameters should be correlated, supposing 
$\boldsymbol{\mu}$ and $\boldsymbol{\sigma}$ are the scenario-specific means and
standard deviations of the LSPs in vector form, and $\mathbf C$ is their correlation matrix, a
sampled vector $\mathbf{x}$ of LSPs is then given as:
\begin{equation}
 \mathbf x=\boldsymbol{\mu}+\boldsymbol{\sigma}\odot\mathbf L\mathbf z,
 \label{eq:lsp_sample}
\end{equation}
where $\mathbf z\sim\mathcal N(\mathbf 0,\mathbf I)$ and $\mathbf L\mathbf L^{\mathsf T}=\mathbf C$. Note that the LSP means, standard deviations, and cross-correlation coefficients in $\boldsymbol{\mu}$, $\boldsymbol{\sigma}$, and $\mathbf C$ can be obtained from the NTN scenario tables in 3GPP TR 38.811, Sec.~6.7.2~\cite{tr38811}.

\subsection{RT-Conditioned Statistical-Cluster Delays Generation}

We then let $L_{\RT}$ be the number of retained RT paths, also known as clusters, $\overline{\tau}_{\RT}$ be their mean excess delay, $N'$ be the number of candidate statistical clusters, and $r_\tau$ be the delay-scaling parameter, where $\overline{\tau}_{\RT}$ is computed from the retained RT paths, $N'$ is the number of statistical clusters to be generated,\footnote{The value of $N'$ can be manually determined or follows the 3GPP suggestion.} 
and $r_\tau$ is the delay-scaling factor specified in 3GPP TR 38.811~\cite{tr38811}.
We first compute the required statistical-cluster mean delay following Step 5 of Sec.~8.4 in 3GPP TR 38.901~\cite{tr38901}:
\begin{equation}
 \mu_{\tau}^{\STAT}=
 \max\!\left[
 r_\tau\mathrm{DS}
 +\frac{L_{\RT}}{N'+1}
 \left(r_\tau\mathrm{DS}-\overline{\tau}_{\RT}\right),
 \overline{\tau}_{\RT}
 \right].
 \label{eq:mu_stat}
\end{equation}
We then let the excess delay of the augmented statistical cluster $n$ be generated using
\begin{equation}
 \widetilde{\tau}_n=-\mu_{\tau}^{\STAT}\ln U_n,
 \label{eq:stat_delay}
\end{equation}
where $U_n\sim\mathcal U(0,1)$. 
To avoid placing a statistical cluster too close to an RT cluster, we remove a candidate cluster if $|\widetilde{\tau}_n-\tau_i^{\RT}|\leq\tau_{\mathrm{th}}$, where $\tau_{\mathrm{th}}=\mu_{\tau}^{\STAT}\ln\!\left(\frac{1}{1-p_0}\right)$. We set $p_0=0.2$, following the example in 3GPP TR 38.901, Sec.~8.4 (Step 5)~\cite{tr38901}.
With the above procedure, we complete the generation of delay for all dominant clusters, regardless of whether they are obtained from the RT simulator or generated statistically.

\subsection{Powers and Angles of Statistical Clusters}

After determining the delay of each statistical cluster, we need to determine their corresponding powers of statistical clusters. To this end, we first generate a virtual power for a cluster $i$, given as:
\begin{equation}
 V_i=\exp\!\left[-\frac{\tau_i(r_\tau-1)}{r_\tau\mathrm{DS}}\right]
 10^{-Z_i/10},
 \label{eq:virtual_power}
\end{equation}
where $Z_i\sim\mathcal N(0,\zeta^2)$. We then let $A=10^{K/10}$ and let $S_V=\sum_i V_i^{\STAT}+\sum_j V_j^{\RT}$, where $V_i^{\STAT}$ is the virtual power of statistical cluster $i$ and $V_j^{\RT}$ is the virtual power of RT cluster $j$. Then, the implemented virtual weighting is given as:
\begin{align}
 P_i^{\STAT,\mathrm{virt}} &= \frac{1}{A+1}\frac{V_i^{\STAT}}{S_V},
 \label{eq:virt_stat}\\
 P_j^{\RT,\mathrm{virt}} &= \frac{1}{A+1}\frac{V_j^{\RT}}{S_V}
 +\frac{A}{A+1}\delta_{j1},
 \label{eq:virt_rt}
\end{align}  
where $\delta_{j1}=1$ only if the RT cluster $j$ is the LoS cluster with smallest delay. Finally, the actual statistical-cluster powers can be determined as:
\begin{equation}
 P_i^{\STAT}=
 \frac{\sum_j P_j^{\RT}}
      {\sum_j P_j^{\RT,\mathrm{virt}}}
 P_i^{\STAT,\mathrm{virt}}.
 \label{eq:power_anchor}
\end{equation}
Note that while we generate virtual powers also for RT clusters, they are only used for computing the powers of statistical clusters, as the RT clusters already have their powers derived from the RT simulator.

After determining the powers of statistical clusters, we then determine the angles. Note that since the procedure of angle generation for azimuth and Zenith AOAs and AODs follows the same principle, here we take azimuth AOA as an example for brevity. To generate the azimuth AOA of statistical cluster $n$, we first compute RT-centered mean azimuth AOA given as:
\begin{equation}
 \phi_{c,\mathrm{AOA}}=
 \arg\!\left(\sum_i P_i^{\RT}e^{j\phi_{i,\mathrm{AOA}}}\right).
 \label{eq:angle_center}
\end{equation}
Then, we let $q_n=\max(P_n^{\STAT}/P_{\mathrm{ref}},10^{-30})$. The azimuth AOA of statistical cluster $n$ is computed as:
\begin{equation}   \phi_{i,\mathrm{AOA}}^{\mathrm{SC}}=\phi_{c,\mathrm{AOA}}+\Delta\phi_n,
\end{equation}
where $\Delta\phi_n=
X_n\frac{2(\mathrm{ASA}/1.4)\sqrt{-\ln q_n}}{C_\phi}+Y_n$, $C_\phi$ is calculated following Eq.~(8.4-11) of 3GPP TR 38.901~\cite{tr38901}, $X_n\in\{-1,+1\}$, and $Y_n$ is a zero-mean Gaussian perturbation. Finally, after all the above steps, we examine whether the generated statistical clusters have powers lower than $25$ dB of the strongest RT cluster. If so, they are again dropped.

\subsection{Sub-Path Generation}
\label{Sec_subpath}

After having the RT clusters and the statistical clusters, we augment subpaths to them, following the procedure in 3GPP TR 38.901~\cite{tr38901}. In the LoS case, the first LoS cluster is represented by a single path, following Step 9 of Sec.~8.4 in 3GPP TR 38.901~\cite{tr38901}, while each remaining cluster is expanded into $M$ subpaths, as specified in 3GPP TR 38.811~\cite{tr38811}.

Under this procedure, all subpaths within the same cluster
share the same delay and are assigned with equal power. Hence, for the $m$-th subpaths of cluster $n$, we have their corresponding delays and powers given by $\tau_{n,m}=\tau_n$ and $P_{n,m}=\frac{P_n}{M_n}$, respectively, where $M_n=1$ for the LoS cluster, otherwise $M_n=M$, and $\tau_n$ and $P_n$ are the delay and power of cluster $n$, respectively.

While the subpaths within a cluster share the same delay and power, their directions are distinguished by introducing intra-cluster angular offsets. Following Step 9 of 3GPP TR 38.901, the AOA, AOD, ZOA, and ZOD of subpath $m$ in each non-LoS cluster $n$ are generated as
\begin{align}
\phi_{n,m,\mathrm{AOA}}
&=\phi_{n,\mathrm{AOA}}
+c_{\mathrm{ASA}}\alpha_m,
\nonumber\\
\phi_{n,m,\mathrm{AOD}}
&=\phi_{n,\mathrm{AOD}}
+c_{\mathrm{ASD}}\alpha_m,
\nonumber\\
\theta_{n,m,\mathrm{ZOA}}
&=\theta_{n,\mathrm{ZOA}}
+c_{\mathrm{ZSA}}\alpha_m,
\nonumber\\
\theta_{n,m,\mathrm{ZOD}}
&=\theta_{n,\mathrm{ZOD}}
+\frac{3}{8}10^{\mu_{\lg\mathrm{ZSD}}}\alpha_m,
\label{eq:ray_angles}
\end{align}
where $\alpha_m$ is the normalized ray angular
offset specified in Table 7.5-3 of~\cite{tr38901}, and $c_{\mathrm{ASA}}$, $c_{\mathrm{ASD}}$,
and $c_{\mathrm{ZSA}}$ are the corresponding
intra-cluster angular spreads defined in~\cite{tr38901}, and  $\mu_{\lg\mathrm{ZSD}}$ is the mean
of the logarithmic ZSD distribution. The resulting zenith arrival angles are mapped to the valid angular range according to the 3GPP procedure. After generating the subpath angles, random subpath coupling
is performed following Step 10 of~\cite{tr38901}, and the complex channel coefficient of each subpath is constructed following Step 13 of \cite{tr38901}. These details are referred to~\cite{tr38901} and are omitted here for brevity. After the aforementioned procedure, the channel
coefficient between transmit antenna element $s$ and receive antenna element $u$ for subpath $m$ in cluster $n$ at each time $t$ can be generated and is denoted as $h_{u,s,n,m}(t)$. Then, the frequency response between
antenna elements $s$ and $u$ is obtained by
summing the contributions of all subpaths:
\begin{equation}
H_{u,s}(\Delta f_k,t)=
\sum_n\sum_{m=1}^{M_n}
h_{u,s,n,m}(t)
e^{-j2\pi\Delta f_k\tau_{n,m}},
\label{eq:cfr}
\end{equation}
where $\Delta f_k$ is the considered baseband frequency
offset from the carrier frequency and
$\tau_{n,m}$ is the propagation delay
of ray $m$ in cluster $n$.
\section{Spatial Consistency for Time-Varying NTN Channels}\label{sec:consistency}
When considering a moving satellite scenario, independently redrawing the statistical channels at
every time step may destroy the temporal continuity. To address this, we therefore implement an explicit frozen-scatterer statistical-consistency module to ensure spatial consistency. Specifically, suppose we want to generate channels for consecutive $T$ time-slots $t_1,t_2,..,t_T$. Then, for RT clusters, they simply follow the RT simulator to maintain the spatial consistency. For the statistically generated clusters, we first generate the statistical channels of $t_1$ following the same procedure as in Sec. \ref{sec:hybrid}. Subsequently, to generate the statistical clusters at $t_2$, we need to use the results of $t_1$ to compute the geometric scatterers of the clusters, and then use their geometry to infer the results of $t_2$. Specifically, we let $d_0$ be the inter-satellite-receiver distance of the first RT LoS path at $t_1$, $\tau_{e,n}$ be the
initial excess delay of statistical cluster $n$, and $\mathbf u_n$ be the unit direction vector for AOA of cluster $n$. The initial total propagation distance of cluster $n$ is $D_{0,n}=d_0+c\tau_{e,n}$. Then, we denote $u_{\mathrm{LoS}}$ as the unit direction vector of the LoS. Then, the angle between the AOA of cluster $n$ and the AOA of the LoS path is $\cos\psi_n=\mathbf u_n\cdot\mathbf u_{\mathrm{LoS}}$. Hence, the inferred
inter-receiver-scatterer distance of path $n$ is
\begin{equation}
 R_{s,n}=\frac{D_{0,n}^2-d_0^2}
 {2(D_{0,n}-d_0\cos\psi_n)}.
\end{equation}
As a result, the scatterer position can be obtained as:
\begin{equation}
 \mathbf s_n=\mathbf p_{\mathrm{Rx}}+R_{s,n}\mathbf u_n.
\end{equation}
Subsequently, we denote $\mathbf p_{\mathrm{sat}}(t_2)$ as the position of the satellite at time $t_2$, the delay of the statistical cluster $n$ can be updated following the geometry as:
\begin{equation}
 \tau_n(t_2)=
 \frac{\|\mathbf p_{\mathrm{sat}}(t_2)-\mathbf s_n\|+R_{s,n}}{c},
 \label{eq:consistent_delay}
\end{equation}
where $\mathbf p_{\mathrm{sat}}(t)$ is the position of the receiver at time $t_2$. Therefore, with this, we use the same scattering point to generate the same cluster to maintain spatial consistency, with delay, power,
departure direction, and propagation evolving with the satellite geometry. Finally, to maintain the consistency of the Doppler term, we let the Doppler shift of cluster $n$ be
\begin{equation}
 f_{D,n}(t)=-f_c\frac{\dd\tau_n(t)}{\dd t}.
\end{equation}
After the parameters of statistical clusters are updated for time $t_2$, we then re-run the sub-path expansion in Sec. \ref{Sec_subpath}, as they are deemed to be small randomness that could still exist even under spatial consistency. We remark that, because this spatial consistency mechanism assumes fixed scattering points for the statistically generated clusters, its accuracy may gradually degrade as the simulation evolves over time. Once the resulting geometry becomes insufficiently representative of the actual propagation environment, the scattering points should be regenerated to refresh the underlying geometric configuration.

\section{Simulation Results and Validation}\label{sec:validation}

This section shows and validates the results through the LSP comparisons and illustration of PDPs.

\subsection{Simulation Setup and Evaluation Metrics}

\begin{figure}[t]
\centering
\includegraphics[width=0.9\columnwidth]{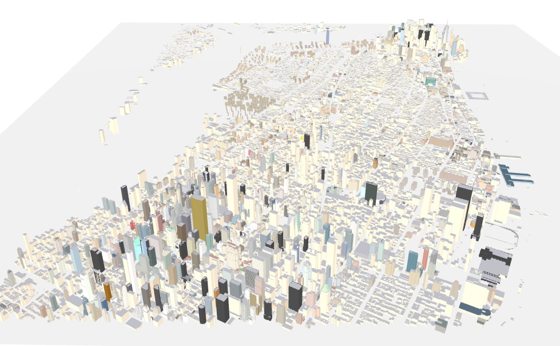}
\vspace{-5pt}
\caption{Manhattan scene used by RT simulations.}
\vspace{-10pt}
\label{fig:scene}
\end{figure}

\begin{figure}[t]
\centering
\includegraphics[width=0.9\columnwidth]{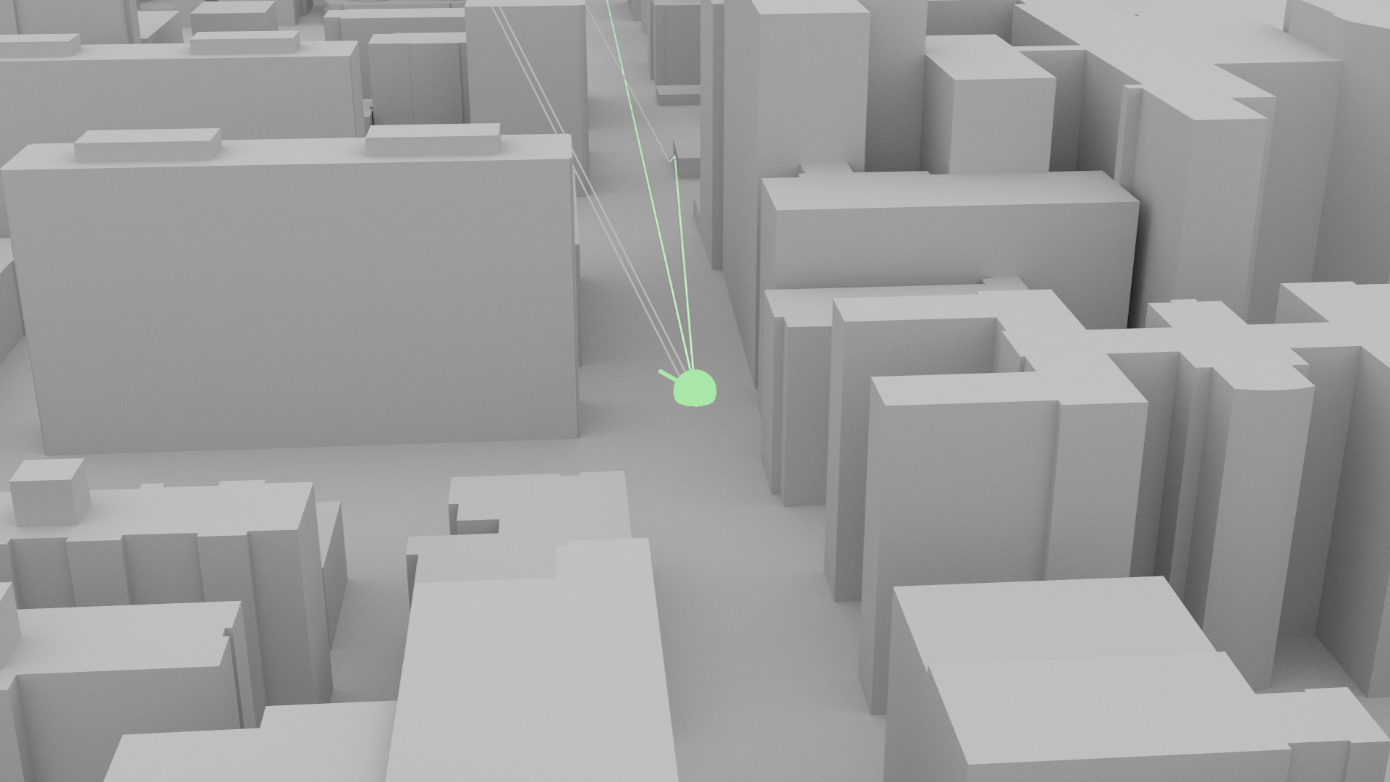}
\vspace{-5pt}
\caption{Illustration of the RT geometry for a receiver.}
\vspace{-10pt}
\label{fig:sparse}
\end{figure}

The configuration adopted for simulation considers a 28-GHz carrier, isotropic one-element RT probes, diffuse scattering with coefficient $0.25$ and a Lambertian pattern, and a full RT simulation with maximum depth of five and $3\times10^6$ rays. The LoS, specular reflection, diffuse reflection, and refraction are enabled for RT, and diffraction is disabled.
The LSP evaluation comprises $50$ receiver positions at $40^\circ$ and $70^\circ$ satellite elevations. The satellite is at a distance of $600$ km. The $0.2$ second, five-sample trajectory configuration is considered for simulations. Table~\ref{tab:defaults} summarizes the simulation parameters adopted in this section. Fig.~\ref{fig:scene} shows the Manhattan scene used for the
RT simulations. Fig.~\ref{fig:sparse} illustrates a representative RT geometry. The 3GPP simulator is based on a dense-urban scenario.

\begin{table}[t]
\centering
\caption{Simulation Configuration}
\label{tab:defaults}
\footnotesize
\begin{tabular}{@{}lr@{}}
\toprule
Parameter & Representative value \\
\midrule
Carrier frequency & 28 GHz \\
Evaluation elevations & $40^\circ$, $70^\circ$ \\
LSP positions per elevation & 50 \\
Trajectory time span / samples & 0.2 s / 5 \\
RT samples per source & $3\times10^6$ \\
Maximum depth & 5 \\
RT pre-filter & $-25$ dB relative \\
Scattering coefficient & 0.25 \\
Scattering pattern & Lambertian \\
RT Tx/Rx probes & $1\times1$ isotropic \\
\bottomrule
\end{tabular}
\end{table}

Suppose the power of a subpath $p$ is $P_p$, the mean delay is derived as $\overline{\tau}=\sum_p w_p\tau_p$, where $w_p=\frac{P_p}{\sum_qP_q}$. Then, the root-mean-square (RMS) delay spread is derived as $\mathrm{DS}=\sqrt{\sum_p w_p(\tau_p-\overline{\tau})^2}$. The Rician K-factor is measured from LoS and NLoS powers, given as $K_{\mathrm{dB}}=10\log_{10}\frac{P_{\mathrm{LoS}}}{P_{\mathrm{NLoS}}}$. For the measure of angle spread, we consider the azimuth AOA as an example, and the same approach applies to other angles. Specifically, we first obtain the mean angle as a power-weighted circular mean angle as $\overline{\phi}_{\text{AOA}}=\arg\!\left(\sum_p w_pe^{j\phi_{p,\text{AOA}}}\right)$, where $\phi_{p,\text{AOA}}$ is the AOA of subpath $p$. Then, the circular RMS AOA spread is derived as $\sigma_{\phi}=\sqrt{\sum_pw_p\,\mathrm{wrap}(\phi_{p,\text{AOA}}-\overline{\phi}_\text{AOA})^2}$.  Following the similar comparison format of the 3GPP tables, the DS, AOA, ADO, ZOA, and ZOD are compared on a log-scale.

\subsection{Simulation Results}

Table~\ref{tab:lsp40} shows the LSP comparison results at $40^\circ$ elevation when using Sionna RT, Hybrid, and the 3GPP simulators.\footnote{The pathloss of RT is also verified for Sionna RT, where the $176.95$ dB is obtained for RT, while the theoretical free-space pathloss is $176.96$ dB.} We see that the delay spread and Rician factor for the hybrid simulator agree well with those of the 3GPP simulator. In addition, while the angle spreads of the hybrid simulator are slightly different than those of the 3GPP simulator because the hybrid simulator is more site-specific, they are still close to each other. On the other hand, we see that the Sionna RT simulator reports somewhat different values of the delay spread and Rician factor. This might be because the rays of the RT simulator cannot appropriately simulate the subpaths under reasonable complexity for the NTN scenarios. Hence, the LoS effect in the RT simulator is somewhat more significant than that of the hybrid and 3GPP simulators. In addition, the angle spreads of the RT simulator are also different than those of the hybrid and 3GPP simulators. Table~\ref{tab:lsp70} reports the measured LSPs at $70^\circ$ elevation. The results again show that the hybrid and 3GPP simulators have closer LSP statistics, and the Sionna RT simulator has different statistics. In addition, we see that when comparing results at $70^\circ$ with those at $40^\circ$, the Rician factor indeed decreases with respect to an increase in the satellite elevation, which agrees with the results observed in the 3GPP channel model. Note that here we only focus on the statistics under the LoS scenario, and the probability of having a LoS scenario indeed decreases with respect to increase of the increase of the satellite elevation. 


\begin{table}[t]
\centering
\caption{LSP Comparison for $40^\circ$ Elevation}
\label{tab:lsp40}
\footnotesize
\setlength{\tabcolsep}{5.0pt}
\begin{tabular}{@{}lrrrrrr@{}}
\toprule
& \multicolumn{2}{c}{Sionna RT} & \multicolumn{2}{c}{Hybrid RT-3GPP}
& \multicolumn{2}{c}{3GPP} \\
\cmidrule(lr){2-3}\cmidrule(lr){4-5}\cmidrule(l){6-7}
Parameter & Mean & Std. & Mean & Std. & Mean & Std. \\
\midrule
$\log_{10}\mathrm{DS}$ & -8.201 & 0.569 & -8.313 & 0.524 & -8.020 & 0.720 \\
$K$ [dB]                & 13.377 & 2.387 & 10.376 & 2.115 & 10.300 & 3.300 \\
$\log_{10}\mathrm{ASA}$ & 0.381 & 0.883 & -0.253 & 2.107 & 0.430 & 0.780 \\
$\log_{10}\mathrm{ASD}$ & -3.429 & 0.635 & -3.270 & 0.989 & -2.810 & 0.340 \\
$\log_{10}\mathrm{ZSA}$ & 1.191 & 0.128 & 1.271 & 0.060 & 1.230 & 0.040 \\
$\log_{10}\mathrm{ZSD}$ & -4.104 & 0.385 & -2.948 & 0.734 & -2.510 & 0.570 \\
\bottomrule
\end{tabular}
\end{table}

\begin{table}[t]
\centering
\caption{LSP Comparison for at $70^\circ$ Elevation}
\label{tab:lsp70}
\footnotesize
\setlength{\tabcolsep}{5.0pt}
\begin{tabular}{@{}lrrrrrr@{}}
\toprule
& \multicolumn{2}{c}{Sionna RT} & \multicolumn{2}{c}{Hybrid RT-3GPP}
& \multicolumn{2}{c}{3GPP} \\
\cmidrule(lr){2-3}\cmidrule(lr){4-5}\cmidrule(l){6-7}
Parameter & Mean & Std. & Mean & Std. & Mean & Std. \\
\midrule
$\log_{10}\mathrm{DS}$ & -7.934 & 0.404 & -8.276 & 0.228 & -8.340 & 0.390 \\
$K$ [dB]                & 8.803 & 0.163 & 7.096 & 1.058 & 8.000 & 1.500 \\
$\log_{10}\mathrm{ASA}$ & 0.588 & 0.773 & -0.011 & 1.228 & 0.180 & 0.640 \\
$\log_{10}\mathrm{ASD}$ & -2.785 & 0.476 & -2.652 & 0.691 & -2.460 & 0.400 \\
$\log_{10}\mathrm{ZSA}$ & 1.649 & 0.001 & 1.682 & 0.025 & 1.650 & 0.070 \\
$\log_{10}\mathrm{ZSD}$ & -3.818 & 0.499 & -2.998 & 0.388 & -2.850 & 0.310 \\
\bottomrule
\end{tabular}
\vspace{2pt}
\end{table}

\begin{figure}[t]
\centering
\begin{minipage}{0.45\textwidth}
\centering
\includegraphics[width=\linewidth]{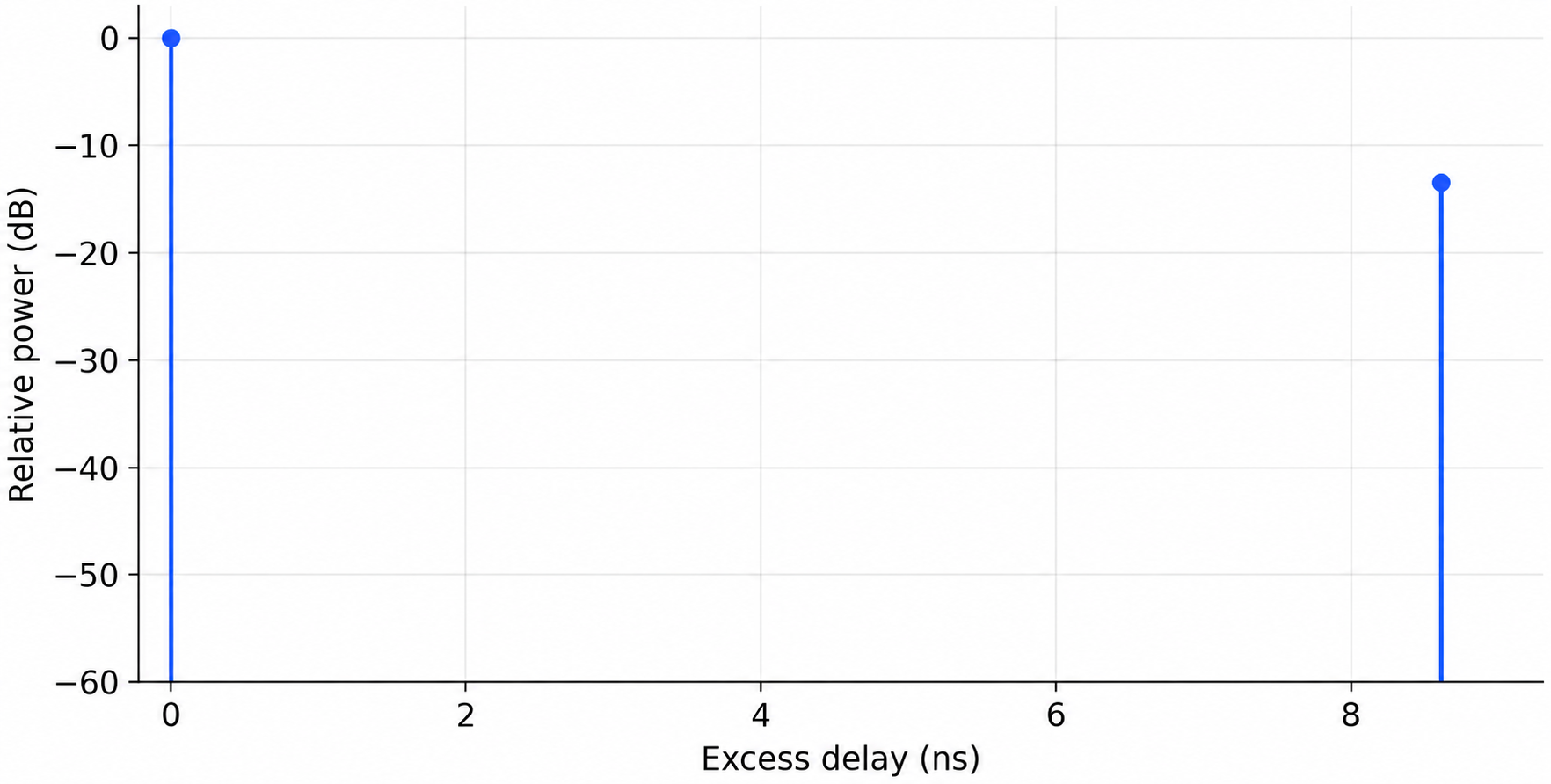}
{\footnotesize (a) Pure-RT (Sionna)}
\end{minipage}
\hfill
\begin{minipage}{0.45\textwidth}
\centering
\includegraphics[width=\textwidth]{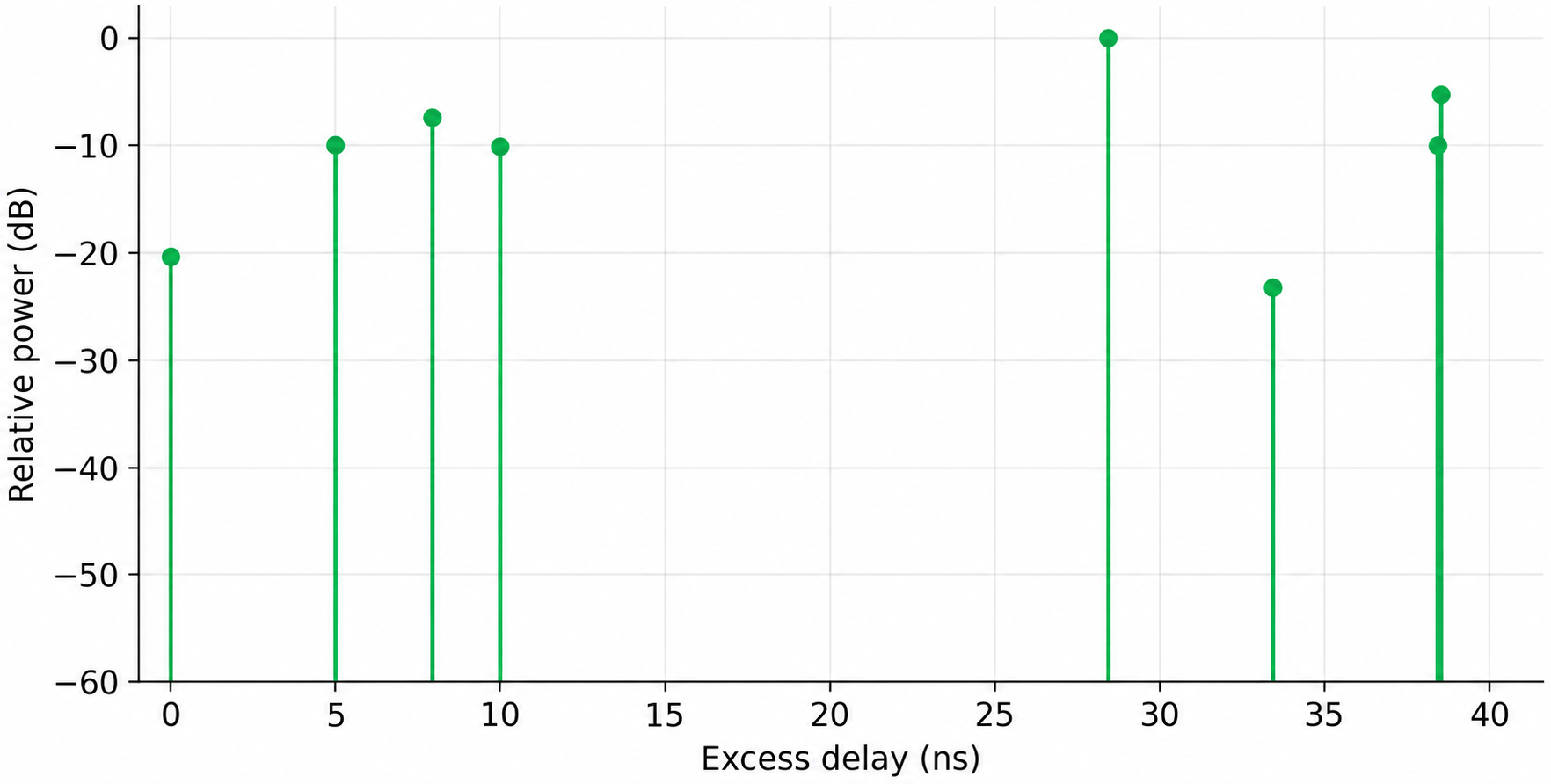}
{\footnotesize (b) 3GPP (OpenNTN)}
\end{minipage}
\par\medskip
\begin{minipage}{0.45\textwidth}
\centering
\includegraphics[width=\linewidth]{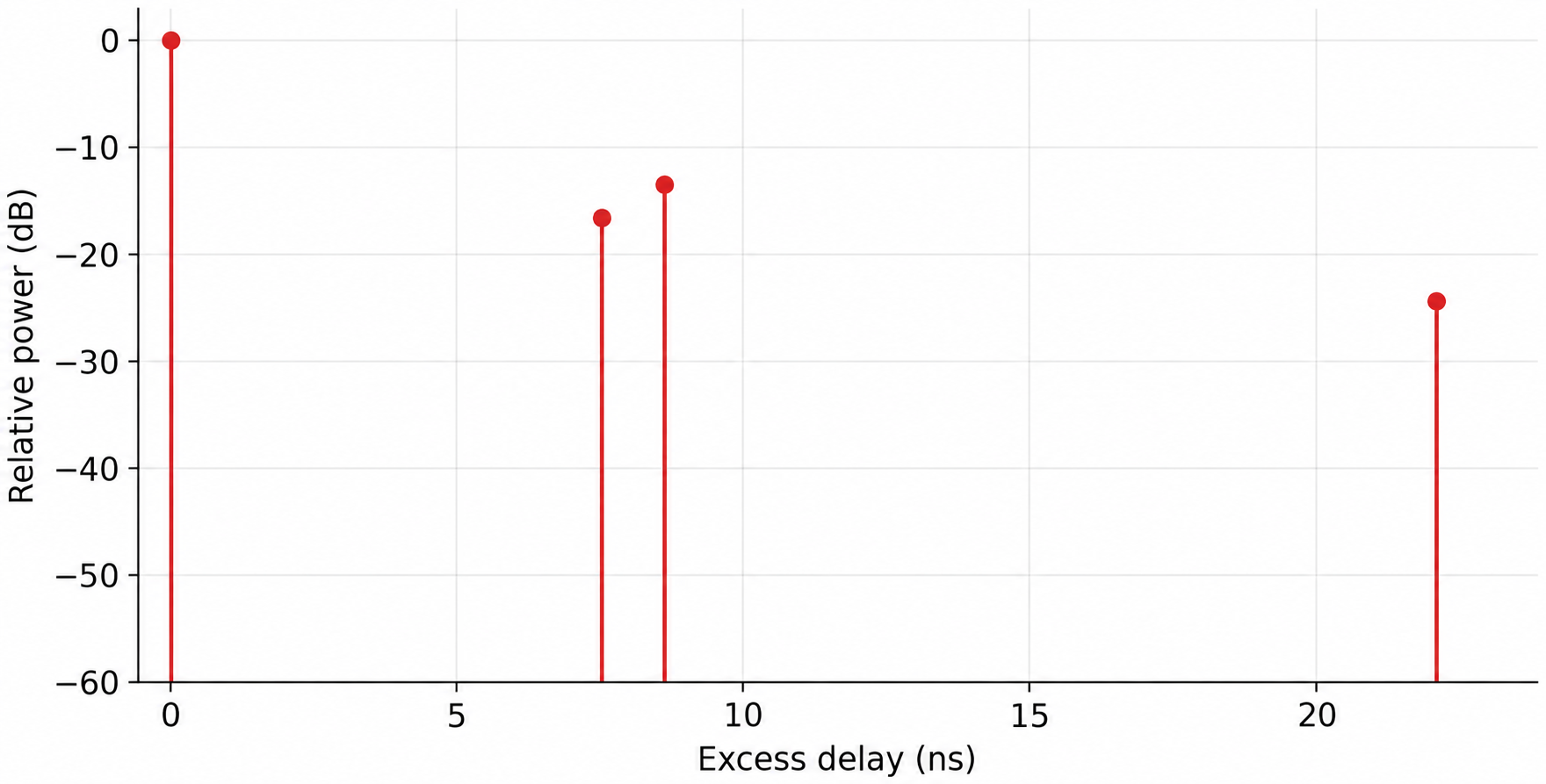}
{\footnotesize (c) Hybrid RT-3GPP (Proposed)}
\end{minipage}
\caption{PDP comparison for a receiver under different channel simulators.}
\label{fig:pdp}
\end{figure}

Fig.~\ref{fig:pdp} compares PDPs of the same receiver under three different channel simulators with satellite elevation of $40^\circ$. We can see that the Sionna RT simulator contains a dominant LoS component and very few additional delayed multipath components, whereas the
3GPP shows a richer multipath structure over a wider
excess-delay range. The hybrid simulator retains the dominant deterministic components while adding statistical multipath components, making it somewhat in between the Sionna RT and 3GPP simulators. Our observations are that the Sionna RT simulator can suffer from too small a number of multipath components, making it less suitable for simulations sensitive to small-scale fading. On the other hand, while the 3GPP simulator can provide rich scattering, it cannot offer site-specific results, and all realizations are based only on statistical modeling. The proposed hybrid simulator therefore combines the advantages of both RT-based and statistical modeling-based simulators by integrating site-specific RT components with statistically generated multipath components, making it particularly suitable for system evaluation and validation requiring the multipath-rich, site-specific condition.

\section{Conclusions}\label{sec:conclusion}

In this paper, we developed a complete NTN channel simulation framework consisting of Sionna RT, 3GPP, and hybrid RT-3GPP simulators. The proposed hybrid approach combines site-specific RT-generated dominant paths with statistically generated paths and subpaths to provide richer multipath realizations under feasible complexity. Simulation results show that the proposed method preserves site-specific propagation characteristics while overcoming the multipath sparsity of the conventional RT-based simulation. The complete simulator has been open-sourced on GitHub.

\bibliographystyle{IEEEtran}
\bibliography{references}

\end{document}